# Quantification of surface displacements and electromechanical phenomena via dynamic atomic force microscopy

Nina Balke[1*], Stephen Jesse[1], Pu Yu[2,3,4], Ben Carmichael[5], Sergei V. Kalinin[1] and Alexander Tselev[1]

[1]Center for Nanophase Materials Sciences, Oak Ridge National Laboratory, Oak Ridge, TN 37831 United States

[2]State Key Laboratory for Low-Dimensional Quantum Physics, Department of Physics, Tsinghua University, Beijing, China

[3]Collaborative Innovation Center of Quantum Matter, Beijing, China

[4]RIKEN Center for Emergent Matter Science (CEMS), Wako, Saitama 351-0198, Japan

[5]Southern Research, Birmingham, AL 35211, United States

*Corresponding author: balken@ornl.gov

## Abstract

Detection of dynamic surface displacements associated with local changes in material strain provides access to a number of phenomena and material properties. Contact resonance-enhanced methods of Atomic Force Microscopy (AFM) have been shown capable of detecting ~1-3 pm-level surface displacements, an approach used in techniques such as piezoresponse force microscopy, atomic force acoustic microscopy, and ultrasonic force microscopy. Here, based on an analytical model of AFM cantilever vibrations, we demonstrate a guideline to quantify surface displacements with a high accuracy by taking into account the cantilever shape at the first resonant contact mode depending on the tip-sample contact stiffness. The approach has been experimentally verified and further developed for the piezoresponse force microscopy (PFM) using well-defined ferroelectric materials. These results open up a way to accurate and precise measurements of the surface displacement as well as piezoelectric constants at the pm-scale with nanometer spatial resolution and will allow avoiding erroneous data interpretations and measurement artefacts. This analysis is directly applicable to all cantilever-resonance-based SPM techniques.

**Introduction**

AFM-based detection of sample surface displacements associated with local changes in the material strain provides access to a number of material properties and phenomena at the nanoscale such as polarization in ferroelectrics [1], thermal expansion due to Joule heating [2], light-matter interaction [3], ion dynamics in ionic conductors [4], mechanical stiffness [5, 6], and others. Quantitative measurements of dynamic surface displacements using an SPM probe can offer important insights into material functionalities at the nanoscale with high a temporal resolution and is desirable for SPM-based characterization techniques such as piezoresponse force microscopy (PFM) [1, 7-11], electrochemical strain microscopy (ESM) [12, 13], atomic force acoustic microscopy (AFAM) [14, 15], or ultrasonic force microscopy (UFM) [16-18]. While these techniques have been developed decades ago, quantification of measured surface displacements is still elusive and would benefit the nanoscale characterization of functional materials. Furthermore, lack of quantitative and accurate measurement can lead to the misinterpretation of relevant material physics. Only if quantitative material parameters can be extracted, can a correlation of nanoscale and macroscale material behavior and structure-function relationships be derived. Many of the above mentioned techniques utilize the dynamic cantilever vibrations when in contact with the sample as a response to an external stimulus. To achieve better sensitivity and signal-to-noise ratio, these and similar techniques use harmonic stimuli, such as voltage bias applied to the tip, sample or probe vibrations, and the AFM photodetector signal is detected at the frequency of the stimulus using a lock-in technique. Resonance enhancement allows further significant improvement in the sensitivity of the measurements by setting the stimulus frequency equal or close to the resonance frequency of one of the cantilever vibrational modes [19, 20]. As a result, quantification efforts become more complex since contact cantilever dynamics need to be taken into account.

In many AFM-based techniques, a conducting AFM tip is placed into contact with a sample surface and a voltage bias is applied between the tip and an electrode beneath the sample. In particular, in electromechanical measurements, the electric field concentrated under the tip apex leads to a change in cantilever vertical position due to a change in the sample volume. The large strength of the highly localized field (on the order of $10^8$-$10^9$ V/m) can also be used as a local tool for material modifications [21-23], or for studies of material behavior in strong electric fields such as nanoscale melting [24] and water condensation [25]. Strong electric fields under the AFM tip can induce water dissociation and water-induced electrochemical reactions [26], material modifications and ionic conduction [27], i.e., the phenomena, which are of a high relevance in nanoscale studies of organic and inorganic material systems. Such experiments require knowledge of the electric field strength and associated strain responses to understand the local functionality.

In turn, resonance enhancement makes the electromechanical techniques highly sensitive, enabling detection of tip displacements in sub-pm range. As a corollary, special attention should be given to possible parasitic effects, which may interfere with the measurements. One such effect is the presence of the electrostatic force acting on the cantilever shank and on the tip apex. The large electric field under a sharp apex of a probe tip can results in a force large enough to produce detectable apex displacements against the mechanical contact stiffness, especially in experiments with a strongly polarizable material, such as a high-permittivity dielectric [28]. The force on the apex is dependent on the local dielectric properties as well as topography of the surface, and can produce local contrasts superimposed onto material electromechanical response of interest. Measurements of the force produced by the electric field may help manage and account for such side effects in the measurements.

Here, we describe in detail the quantification of surface displacements with use of cantilever resonance at the first flexural contact resonant mode for the example of PFM measurements on ferroelectric materials. From this, material properties, such as the piezoelectric constant $d$, can be extracted. Recently, first steps towards quantification of $d$ have been made using an interferometric approach [29], which requires additional hardware and until now it cannot be done with any commercial ambient SPM. Our goal is to provide a pathway to parameter quantification which is based on pure data analytics. For this, the cantilever mode shapes are calculated using an analytical model and the cantilever response functions is extracted for a variety of cantilevers. We show that quantification of the surface and tip apex displacements can be achieved by performing a static calibration of cantilever sensitivity and application of a correction factor to account for the cantilever mode shape depending on the cantilever and the tip-sample contact properties. In order to verify our approach, we compare theoretical data with PFM measurements using a variety of cantilevers with a range of properties on high-quality well-characterized ferroelectric materials. We extract the quantitative piezoelectric constant for a commercial periodically poled lithium niobate (PPLN) crystal and find an excellent match with a known value specified by the sample manufacturer. In addition, the ferroelectric switching process is evaluated for an ferroelectric epitaxial film of tetragonal $Pb(Zr,Ti)O_3/(La,Sr)MnO_3$ (PZT/LSMO). All measurements are done in a controlled low-humidity environment (Ar-filled glove box). In the analysis, we are able to elucidate the influence of the cantilever properties and explain the measured scatter in PFM values obtained with different cantilevers.

**Resonance-enhanced dynamic cantilever displacements**

We begin with description of a general approach currently used for semi-quantitative cantilever displacement measurements. The amplitude of the AFM photo-detector signal $V^{ph}(\omega)$ produced by a harmonic displacement $u(t) = u_0 \sin(\omega t)$ of the sample surface in contact with the tip can be expressed as:

$$V^{ph}(\omega) = s\theta(\omega) = sS(\omega)u_0 \qquad (1)$$

where $s$ is the conversion factor between the change of the cantilever slope $\theta$ at the position of the sensing laser beam and the photo-detector output voltage, and $S(\omega)$ is the mechanical transfer function of the cantilever. The response of the cantilever $S(\omega)$ to a small harmonic excitation with an angular frequency ω close to an eigenfrequency of a resonant mode $\omega_0$ can be well described as a response of a simple harmonic oscillator (SHO) with an oscillation amplitude [30, 31]:

$$|H(\omega)| = \frac{H_0 \omega_0^2}{\sqrt{(\omega^2 - \omega_0^2)^2 + (\omega \omega_0/Q)^2}} \qquad (2)$$

where, $Q$ is the quality factor of the mode (typically on the order of 100 for the lowest flexural contact mode), and the factor $H_0$ is directly proportional to the amplitude of the driving excitation. At resonance, $|H(\omega_0)| = H_0 \cdot Q$, and the slope amplitude $\theta(\omega_0) = s'u_0 \cdot Q$. Correspondingly, for the amplitude of the photodetector signal:

$$V^{ph}(\omega_0) = s \cdot s' \cdot u_0 \cdot Q \qquad (3)$$

The use of multifrequency measurements, such as Band Excitation (BE) or Dual AC Resonance Tracking (DART) [9, 30], allows for quantitative characterization of the cantilever response at a resonance. It allows precise measurements of the contact resonance frequency, Q-factor, and the amplitude $V^{ph}(\omega_0)$ at each pixel of the image. In particular, in BE, the measured contact resonance peak is fitted using eq. 2, which provides the PFM amplitude in units of Volts of the photodetector signal. However, conversion of the PFM signal into parameters of the material response remains highly non-trivial. In particular, quantification of the cantilever displacement $u_0$ requires appropriate calibration of the $s \cdot s'$ product in eq. 3, and it is usually performed by measuring the cantilever sensitivity by static force-distance curves, in which case:

$$V_{\text{static}}^{ph} = s \cdot \theta_{\text{static}} = s \cdot s'_{\text{static}} \cdot u_0. \qquad (4)$$

A complication arises because generally, static and dynamic sensitivities of a cantilever are not equal: $s' \neq s'_{\text{static}}$. First, we note that during the static calibration, cantilever bending is much larger than in dynamic vibrations during dynamic detection techniques, and the force cannot prevent the slippage of the tip apex long the surface (bonded vs. sliding indentation). Therefore, in the static calibrations, the cantilever shape is imposed exclusively by the vertical displacement of the cantilever free end in respect to the cantilever root (clamped end) equivalent to the displacement of the sample along surface normal. In the static calibration, the distance of the tip apex slip along the surface is determined by the deformed cantilever shape and not by the force at the tip-sample contact.

The stiffness of the tip-surface junction will play a significant role in the relevant dynamics, since it determines corresponding boundary conditions for the beam. However, the contact stiffness is position dependent and is determined by local topography and indentation modulus of material [32-36]. Consequently, its effect cannot be inferred beforehand due to a lack of precise information about the cantilevers and/or the sample properties.

Below, we analyze in detail the cantilever vibration motion in response to the periodic vertical surface displacement. The analysis is based on an idealized cantilever model shown in figure 1a. The model is a beam of rectangular cross-section and a length $L$ rigidly clamped at one end. The beam tiled in respect to the sample surface at an angle $\varphi$. A massless and rigid sensing tip of a length $h$ is located at a distance $L_2$ from the beam free end and makes a right angle with the beam. Boundary conditions at the tip-sample contact are represented by two Kelvin–Voigt linear elements with springs accounting for the contact stiffness $k^*$, $k^*_{Lon}$ and a dashpots describing contact damping $\gamma$, $\gamma_{Lon}$ in directions normal to and along the surface, respectively. This idealized model takes into account both normal and longitudinal contact stiffness contact damping and cantilever tilt. It, however, ignores the finite mass of the sensing pyramid, possible flexural deformations at its apex, as well as finite dimensions of its base. It also ignores possible deviations of the cantilever geometry from ideal specifications.

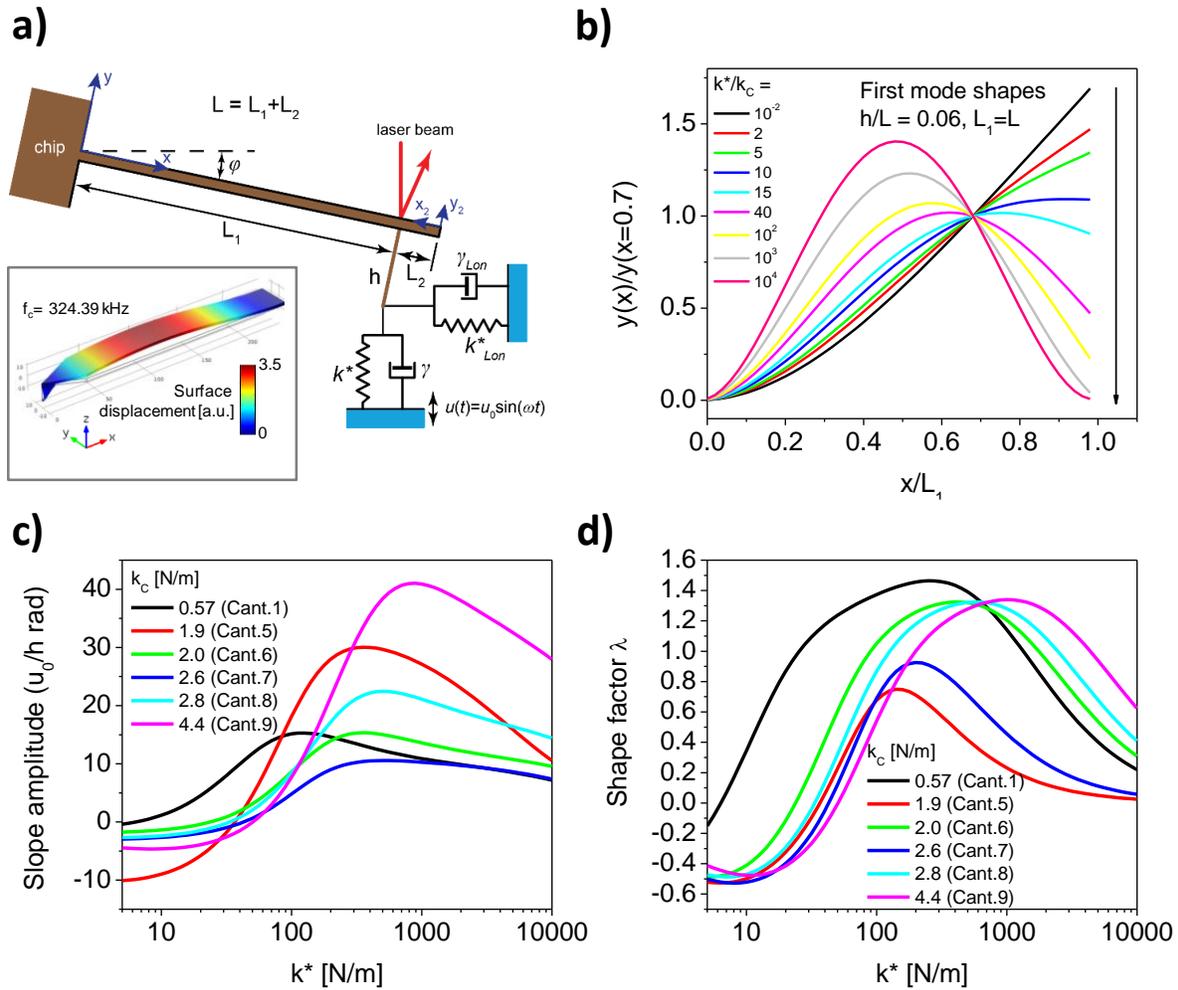

**Figure 1.** (a) Cantilever model used in the analytical calculations (adapted from ref. [37]). See text for explanation of notation. The inlet shows the shape of the first contact flexural mode of an Olympus AC240 cantilever calculated using finite elements modeling with COMSOL v.4.4 software (COMSOL AB). (b) Calculated cantilever shapes for several values of normalized contact stiffness $k^*/k_C$, where $k_C$ is the cantilever spring constant; calculations were performed using the idealized model in panel (a). (c) Calculated cantilever slope amplitudes at resonance at the location of the sensing tip (where the laser spot is typically placed in experiments) as functions of contact stiffness for different cantilevers. Positive and negative values reflect difference in slope sign for a given phase of sample surface oscillation. (d) Shape correction factors as functions of contact stiffness. Cantilever properties are provided in Table S2 of Supplementary data.

The problem for the forced harmonic oscillations of the cantilever model in figure 1a for excitation through the vertical displacement of the cantilever was solved analytically, and the solution is provided in ref. [37]. We use this solution further in our analysis. Here, the analysis is restricted solely to the cantilever vibrations in the linear regime at the first flexural contact eigenmode at resonance excited by harmonic vertical displacements of the sample surface $u(t) = u_0 \exp(i\omega t)$. Further, it is assumed that the Herztian contact mechanics model is valid with $k^*_{Lon}/k^* = \gamma_{Lon}/\gamma = 0.85$ [37, 38]. The cantilever tilt angle was set to $\varphi = 12°$ in the calculations; cantilever dimensions $L$, $L_2$, and $h$ were used as provided by manufacturers.

In AFM-based characterization techniques using dynamic cantilever excitation, the amplitude of the dynamically changing slope at the end of the cantilever is proportional to the measured signal. The latter depends strongly on the contact stiffness $k^*$. Figure 1b shows a series of curves representing first resonant mode shapes for different contact stiffness $k^*/k_C$ calculated with use of the analytical model. Here, $k_C$ is the static spring constant of the cantilever. As seen, with $k^*/k_C$ increasing from a small value, the slope at the free end of the beam changes sign crossing zero at about $k^*/k_C = 10$, increases, but then decreases with its shape approaching that at the clamped end for $k^*/k_C$ approaching infinity. This trend is an effect of the longitudinal component of contact stiffness, which cannot be ignored in the experiments with relatively hard materials. The inset in figure1a illustrates the shape of the first contact eigenmode as a three-dimensional cantilever model for a case of a relatively stiff contact, when the maximum of the displacement amplitude in the middle along the cantilever length is pronounced.

The parameters $k^*$ and $\gamma$ can be found from the experimentally measured contact resonance frequency and Q-factor. To do so, we introduce a complex contact stiffness $\hat{k}^* = k^{*\prime} + ik^{*\prime\prime} = k^* + i(\omega\gamma)$ and complex frequency $\hat{\omega} = \omega' + i\omega''$. Q-factor of the resonance can be expressed through the real and imaginary parts of the complex frequency as $Q = \omega'/(2\omega'')$. Here, $\omega'$ has the meaning of the damped natural frequency of the contact resonance mode. For a strongly underdamped oscillator, $\omega'$ is nearly equal to the resonant frequency at forced oscillations, and we set $\hat{\omega}_{1,c} = \omega_{1,c}(1 + i/2Q_{1,c})$, where $\omega_{1,c}$ and $Q_{1,c}$ are the experimentally measured angular resonant frequency and Q-factor of the first flexural contact resonant mode, respectively. The value of $\hat{\omega}_{1,c}$ is substituted for frequency in the characteristic equation of the cantilever beam $N(\hat{\omega}, \hat{k}^*) = 0$ (eq. S.8 in Supplementary data), and the equation is numerically solved in respect to $\hat{k}^*$. The expressions for the characteristic equations are very lengthy to fit in the format of the paper and provided in Supplementary data. From the solution, the experimental contact stiffness $k^*$ and the damping $\gamma$ are: $k^* = \mathrm{Re}\,\hat{k}^*$ and $\gamma = \mathrm{Im}\,\hat{k}^*/\omega_{1,c}$.

The characteristic equation can be further solved in respect to complex frequency for varying $k^*$ to find resonant frequencies $\omega_{1,c}$, slope amplitudes $A_{\theta1,c}(k^*)$, resonance Q-factors $Q_{1,c}(k^*)$, and shape factors $\lambda(k^*)$ as functions of $k^*$ for a variety of cantilevers. However, first, we need to make a choice for the model of contact damping. In the calculation of $A_{\theta1,c}(k^*)$ and $Q_{1,c}(k^*)$ below, we applied the model used by Yuya et al. [39] in measurements of storage and loss moduli of viscoelastic materials with contact-resonance AFM. For the purpose of our calculations, only the ratio of the contact damping parameter $\gamma$ and contact stiffness $k^*$ are of importance. Yuya et al., who used Hertzian model for contact mechanics, elastic solution for the material Young modulus from nanoindentation, and elastic-viscoelastic correspondence principle, arrived at expressions, which yield:

$$\frac{\omega\gamma}{k^*} = \frac{E''^*}{E'^*} \tag{5}$$

where $E'^*$ and $E''^*$ are reduced storage and loss moduli, respectively. Following Yuya et al. [39], we assumed that both $E'^*$ and $E''^*$ are material properties independent on details of the contact geometry and frequency in the range of typical contact resonance experiments. Therefore, $\omega\gamma/k^* = const$, or $\omega\gamma = \alpha k^*$, where $\alpha$ is a constant. This model describes a situation when energy dissipation in a dynamic process at the contact is proportional to the strain. The coefficient $\alpha$ can be found in this case from experimental data and solution of the characteristic equation $N(\widehat{\omega}, \widehat{k}^*) = 0$ as $\alpha = \operatorname{Im}\widehat{k}^*/\operatorname{Re}\widehat{k}^*$. Alternatively, one may consider a model, when dissipation is proportional to the rate of the strain change. In this case, $\gamma/k^* = const$, and $\gamma = \alpha k^*$ with $\alpha = \operatorname{Im}\widehat{k}^*/(\omega \cdot \operatorname{Re}\widehat{k}^*)$.

After that, the amplitude of the cantilever slope at the position of the sensing tip $A_{\theta1,c}(k^*)$ is calculated from the full analytical solution for the cantilever vibrations using the obtained values, and shape factors $\lambda$ are determined as:

$$\lambda = s'/s'_{\text{static}} = A_{\theta1,c}(k^*)/Q_{1,c}(k^*)/\theta_{\text{static}}. \tag{6}$$

The expression for the static slope $\theta_{\text{static}}$ corresponding to a static cantilever displacement equal to $u_0$ can be readily derived from expression given by Mazeran and Loubet [38] for the case of sliding cantilever and neglecting correction from the friction force:

$$\theta_{static} = \frac{3}{2} \frac{u_0}{L_1 \cos\varphi} \ . \tag{7}$$

Figure 1c shows $A_{\theta 1,c}(k^*)$ as functions of $k^*$ calculated for a broad range of contact stiffness values from 5 N/m to $10^4$ N/m and cantilever geometries (see Table S2 in Supplementary data) for the damping model $\omega\gamma = \alpha k^*$. Figure 1d displays the corresponding shape factors $\lambda(k^*)$ for all cantilevers. As seen, the $A_{\theta 1,c}(k^*)$ curves for all cantilever show a maximum over $k^*$, which is an effect of the longitudinal contact stiffness (similar to that described in ref. [38] for the quasi-static case). The exact shape of the curves and positions of maxima are strongly cantilever-dependent. The $Q_{1,c}(k^*)$ curves reflect a non-monotonic behavior of the Q-factor vs. contact stiffness. As discussed in detail in Ref. [40], this is a result of mode-shape-dependent motion amplitude, and hence, dissipation, of the tip apex at contact. With increasing contact stiffness, the tip-apex motion gradually ceases, and all Q-factors grow.

The choice of the contact dissipation model depends on properties of the specific material. However, importantly, the calculations show that the shape factors as functions of contact stiffness are nearly insensitive to the exact dissipation mechanism. In fact, calculations reveal that the shape factors change by less than 0.1% with change of the model for the contact damping from $\omega\gamma = \alpha k^*$ to $\gamma = \alpha k^*$ at $k^* > 100$ N/m. The shape factors are also very tolerant to the changes of the values of $\gamma$, reflecting a weak dependence of the mode shape on the contact damping. For the contact stiffness range covered by the calculations, the shape factors strongly deviate from unity varying between 0 and 1.5, and therefore it is important that knowledge of the shape is accessed for accurate measurements of the sample surface displacements.

Note that for certain cantilevers—1, 6(8), and 9—the shape factors remains within 30% of unity in a broad range of the contact stiffness, and therefore, these cantilevers should be preferred in the measurements. Additionally, cantilever 9 will yield the largest slope amplitude at the resonance (as can be concluded from figure 2a) setting this tip model apart from others in terms of measurement sensitivity. Such knowledge can be used to compare data measured with different probes and to make an intelligent selection of cantilevers to be able to measure the cantilever displacement quantitatively.

**Piezoelectric constant of a periodically poled lithium niobate crystal**

In order to verify the model developed in the previous section, we apply the data correction to PFM measurements of a piezoelectric constant of a PPLN crystal (Bruker). In PFM, the dynamic cantilever displacement $D_{ac}$ is caused by a local sample volume expansion resulting from the inverse piezoelectric effect and is proportional to the piezoelectric constant $d$. In our case, $D_{ac}$ is synonymous with the so-called PFM signal. Since $d$ is a material property, the cantilever displacement should be independent of the cantilever properties.

The PFM experiments were carried out using the band excitation (BE) technique [30, 34]. The use of the BE technique is crucial here since the model described above requires inputs of a contact resonance frequency as well as a Q-factor, which cannot be extracted from single-frequency measurements. During BE, the full contact resonance peak is fitted with the SHO model eq. 2 to extract response amplitudes and resonance quality factors. All the measurements were performed in an Ar-filled glove box to minimize the effect of the surface water layer and to keep environmental conditions constant for all measurements. The $H_2O$ level in the glove box was <0.1 ppm. Two varieties of metal-coated cantilever probes with stiffness values 0.6 N/m and 3.9 N/m were used (MikroMash and Nanosensor). Cantilever properties are listed in Table S1 of Supplementary data. All the probes were calibrated through force-distance curves and thermal tunes to extract the static cantilever sensitivity and spring constant (stiffness). During the PFM measurements, the contact force was kept at about 120 nN for all the probes.

Before performing quantitative measurements, it is necessary to understand the main PFM signal components and image contrast mechanisms. This becomes obvious from the PFM images taken on the PPLN sample (figure 2a), which show strong asymmetry in signal strength between ferroelectric domains of opposite orientations (blue and red areas in figure 2a) as seen from the histogram in the inset in figure 2a. Since no contrast was observed in the corresponding images of the contact resonance frequency and quality factor, this asymmetry will remain after data quantification.

Recently, we have shown that electrostatic tip-sample interactions can strongly contribute to the measured cantilever displacements [41, 42]. The electrostatic force $F$ is proportional to the applied voltage squared: $F = \frac{1}{2} C' \cdot V^2$. In the electrostatic force microscopy (EFM) and traditional non-contact Kelvin probe force microscopy (KPFM), $C'$ is typically negative and equal to the derivative of the capacitance of the probe as a function of the probe-sample distance $z$: $dC/dz$. In contact, $dC/dz$ becomes undefined, and $C'$ has to be viewed as a proportionality coefficient between the force and voltage with local and global contributions from the tip apex, tip cone, and cantilever shank [43, 44]. With a sinusoidal voltage $V_{ac}$ superimposed on a dc bias $V_{dc}$, $V = V_{dc} + V_{ac} \cdot \sin(\omega t)$, the amplitude of the first harmonic of the electrostatic force $F_{ac}$ can be written down as: $F_{ac} = C' \cdot V_{ac} \cdot V_{dc}$, or $F_{ac} = C' \cdot V_{ac} (V_{dc} - V_{SP})$ in the presence of a surface charge, with $V_{SP}$ being the corresponding surface potential. In the case of PPLN, a measurable surface potential is always present as seen in the image in figure 1b, which was recorded using the peak force (PF) KPFM. This fact needs to be considered in the further data analysis. Depending on the contact stiffness $k^*$, an electrostatic force can result in a measurable cantilever displacement $D_{ac} = F_{ac}/k^*$ with a linear dependence $D_{ac}$ on $V_{dc}$. In turn, for the piezoelectric contribution, $D_{ac}$ is independent of $V_{dc}$ unless domain switching takes place: $D_{ac} = d \cdot V_{ac}$. Both contributions are schematically shown in figure 2c where domains with opposite polarization orientations correspond to positive and negative $D_{ac}$.

In the following, we denote the cantilever displacements due to piezoelectricity and electrostatics as $D_{PE}$ and $D_{ES}$, respectively. Figure 2d schematically illustrates the combined effect of piezoelectric and electrostatic contributions. The original symmetric piezoelectric contribution +/-$D_{PE}$ is shifted by $D_{ES}$, resulting in unequal values of cantilever displacements $D_{ac,1}$ and $D_{ac,2}$ for domains pointing up and down, respectively. This directly affects the PFM imaging, which is performed at 0 $V_{dc}$. In a range of $V_{SP}$ and the slope $m$ of the dependency $D_{ac}$ on $V_{dc}$, the asymmetry can be so strong that no phase flip is observed between different domains.

In order to distinguish between piezoelectric and electrostatic contributions, we apply a combination of KPFM and on-field switching-spectroscopy PFM (SS-PFM) [45] in two sequential measurements on the same area across a sparse grid. During the on-field PFM voltage spectroscopy, $V_{dc}$ and $V_{ac}$ are applied simultaneously and piezoelectric as well as electrostatic signal contribution are measured. After that, the open-loop KPFM is performed in the non-contact mode with the same voltage sequence as used for the on-field PFM measurements. The KPFM provides the distribution of $V_{SP}$, while the on-field PFM voltage spectroscopy yields $m$ as well as $D_{ac,1}$ and $D_{ac,2}$ over domains with opposite orientations. Combining this information, +/-$D_{PE}$ can be calculated. Noteworthy, $V_{dc}$ applied to measure the on-field PFM curves has to be small enough to avoid injecting charges from the tip and altering the domain structure underneath it. Both can be challenging for ferroelectric thin films. In contrast, for PPLN crystals, the coercive voltages are high, and only charge injection needs to be considered. Here, it has been found that fresh tips can inject some charge into the PPLN when a tip is still sharp. After a few measurements, charge injection typically becomes negligible, which is attributed to the tip wear, increasing the tip radius and, thus, decreasing the local electric field driving the charge injection.

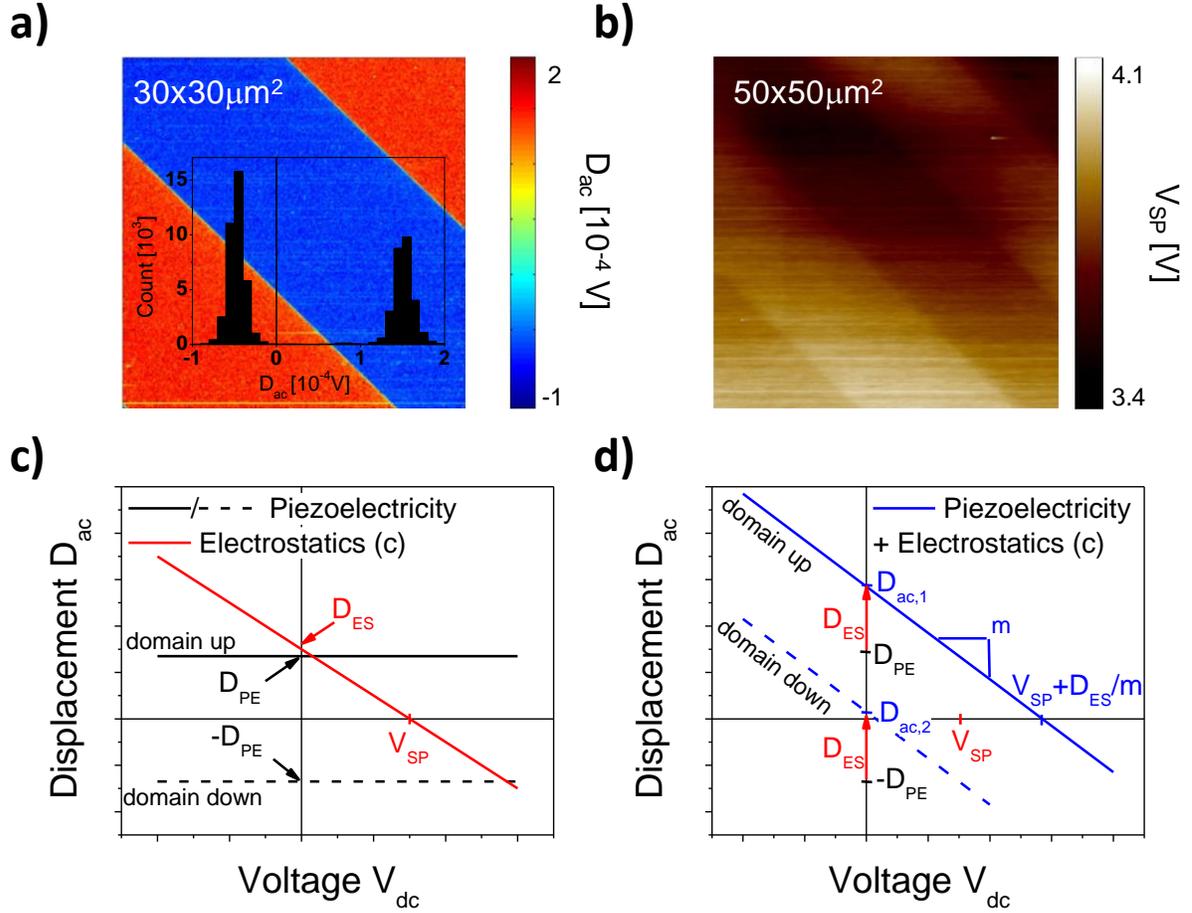

**Figure 2.** (a) A PFM image of domains on the PPLN crystal obtained with $V_{ac} = 2$ $V_{pp}$ using BE before any calibration/quantification. The inlet shows a histogram of the data points in the image. (b) Surface potential image obtained with PF-KPFM on the PPLN crystal. (c) and (d) Schematic analysis of piezoelectric and electrostatic signal contributions considered separated (c) and combined (d).

Measurements on the PPLN crystal were performed with two different cantilevers, A and B, of different stiffness as summarized in Table S1 of Supplementary data. Figure 3a shows the on-field PFM signal plots for the two cantilevers measured on a 128x64 grid over a 50 μm x 50 μm area and averaged over oppositely oriented domains before any data quantification. The plots are straight lines without any hysteresis (compare with figure 2d). As expected, the softer cantilever shows a larger slope *m* corresponding to a higher sensitivity to the electrostatic forces. Based on the cantilever geometry, free as well as contact resonance frequencies and *Q* factors, contact stiffness *k\** and shape factors $\lambda$ were calculated for each point of the grid using the model of the previous section. The resulting histograms are shown in figure 3b and c, respectively.

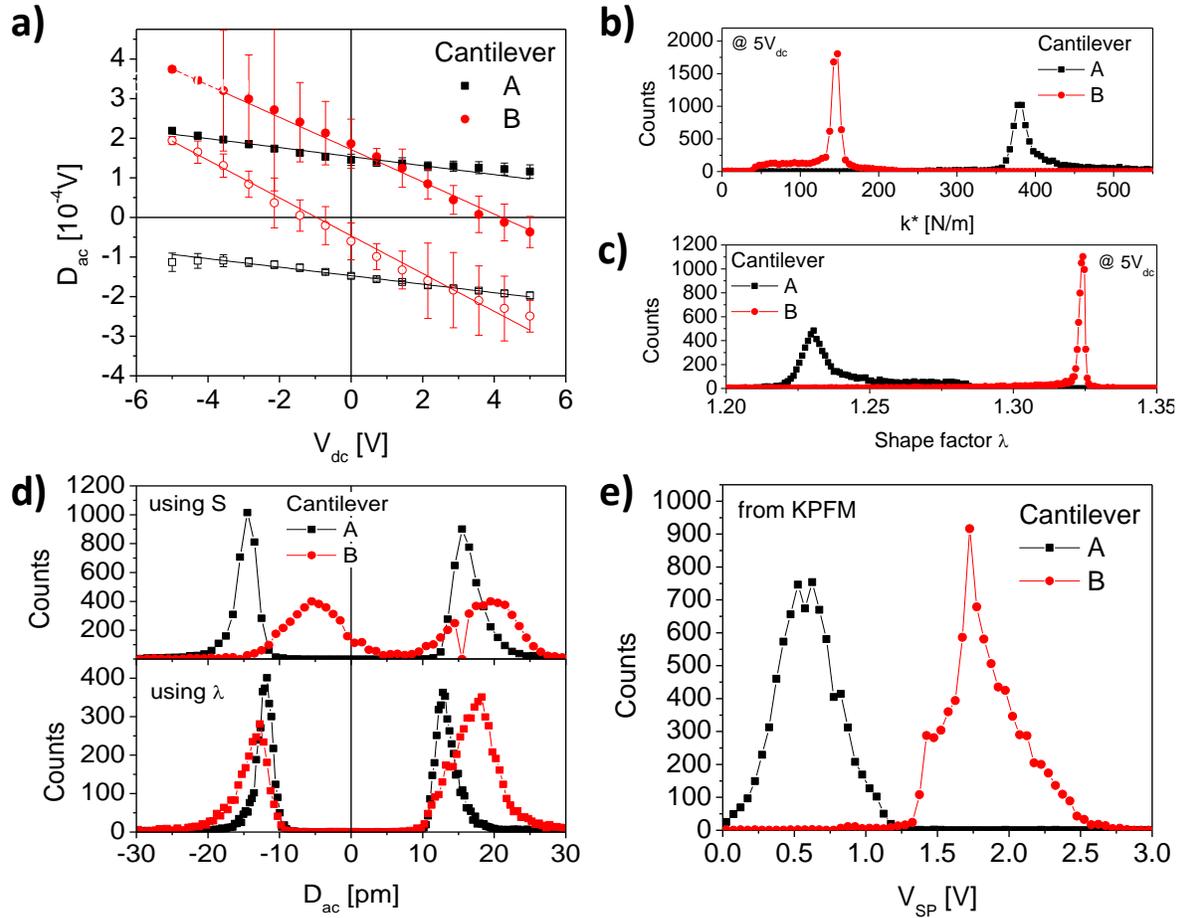

**Figure 3.** (a) PFM signal $D_{ac}$ as a function of $V_{dc}$ measured on a 128x64 point grid over a 50 μm x 50 μm area with two different cantilevers, A and B, and averaged over oppositely oriented domains. Histograms of (b) contact stiffness $k^*$ and (c) shape factor $\lambda$ measured in the same area for cantilevers A and B. (d) Histograms of $D_{ac}$ expressed in picometers with use of the static sensitivity $S$ (top) and the shape factor $\lambda$ (bottom). (e) Histograms of the measured surface potential $V_{SP}$ for two areas over the PPNL crystal.

A typical way of transforming the units of $D_{ac}$ from volts (as measured at the photodetector) to picometers is to use an (inverse) static cantilever sensitivity $S = (s \cdot s'_{\text{static}})^{-1}$ as described at the beginning of the previous section (eq. 4). Figure 3d compares this approach versus our model incorporating the resonant mode shape through the factor $\lambda$. It can be seen that using the static sensitivity only, the extracted values overestimate the surface displacement (i. e., $\lambda > 1$). The distributions of the values for cantilever B are noticeably different between the two methods, showing that it is necessary to apply the cantilever shape correction. Still, despite the improved data quantification, different cantilevers yield significantly different values of $D_{ac}$. This can be explained by variations of the surface potential $V_{SP}$ revealed by the histograms in Figure 3e. The alterations of the surface potential can be associated with the contact electrification that occurs during scanning in the contact mode [42, 46-48]. However, $V_{SP}$ changes of about +/-1V were also observed between different days, which might be related to unknown environmental factors. These observations demonstrate the importance of combining the PFM measurements with the surface potential characterization.

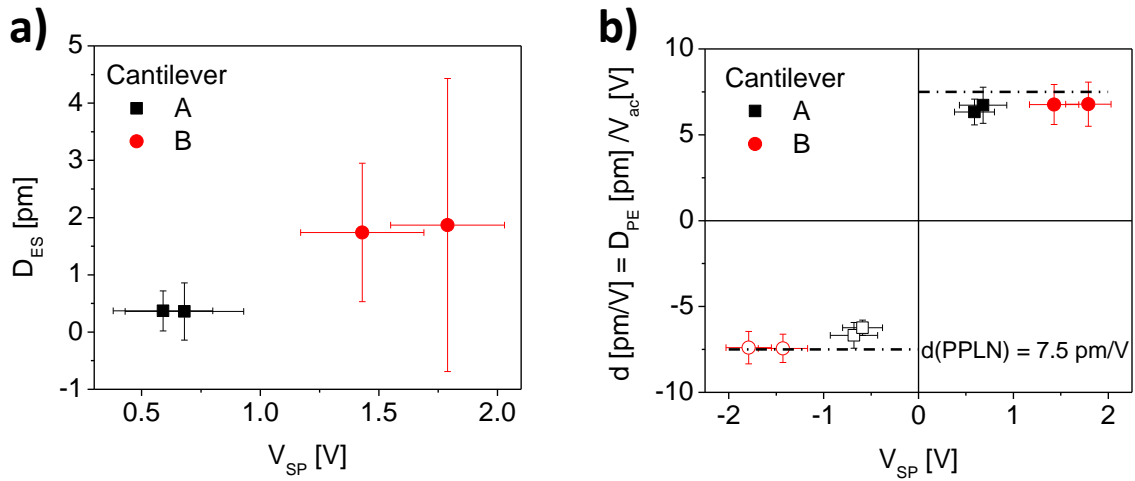

**Figure 4.** Separation of the PFM signal $D_{ac}$ into (a) electrostatic, $D_{ES}$, and (b) piezoelectric, $D_{PE}$, contributions plotted as functions of the measured PPLN surface potential $V_{SP}$ for cantilevers A and B. For comparison, panel (b) also shows the value of the piezoelectric constant provided by the PPLN manufacturer.

After quantification of the shape factor $\lambda$, $V_{SP}$ and $m$, the contributions from electrostatics, $D_{ES}$, and piezoelectricity, $D_{PE}$, can be separated and plotted, as displayed in figure 4. These measurements were carried out in two different areas for each of the two probes. Figure 4a shows the electrostatic contribution. As seen, $D_{ES}$ (at $V_{dc} = 0$ V and $V_{ac} = 2\ V_{pp}$) can be between 0.5 and 2 pm depending on the surface potential. Further, we compare $D_{PE}$ with the piezoelectric constant value $d_{PPLN} = 7.5$ pm/V specified by the PPNL crystal provider. In figure 4b, the measured piezoelectric constants are displayed together with the value 7.5 pm/V. It can be seen that both the cantilevers yield values approaching the manufacturer-specified macroscopic one for domains of both orientations, and the results are independent of the surface potential. This shows the potential for PFM to become a quantitative characterization technique when all signal-contributing mechanism and the cantilever dynamics are considered. It also demonstrates a problem arising for weak ferroelectric materials with low piezoelectric constants. For $V_{ac}$ = 2 $V_{pp}$, the electrostatic contribution was evaluated to be between 0.5 and 2 pm. If a piezoelectric constant would be one order of magnitude lower, for example, the piezoelectric contribution would be in the same range as the electrostatic contribution and, hence, would be difficult to detect.

**Switching properties of an epitaxial PZT thin film**

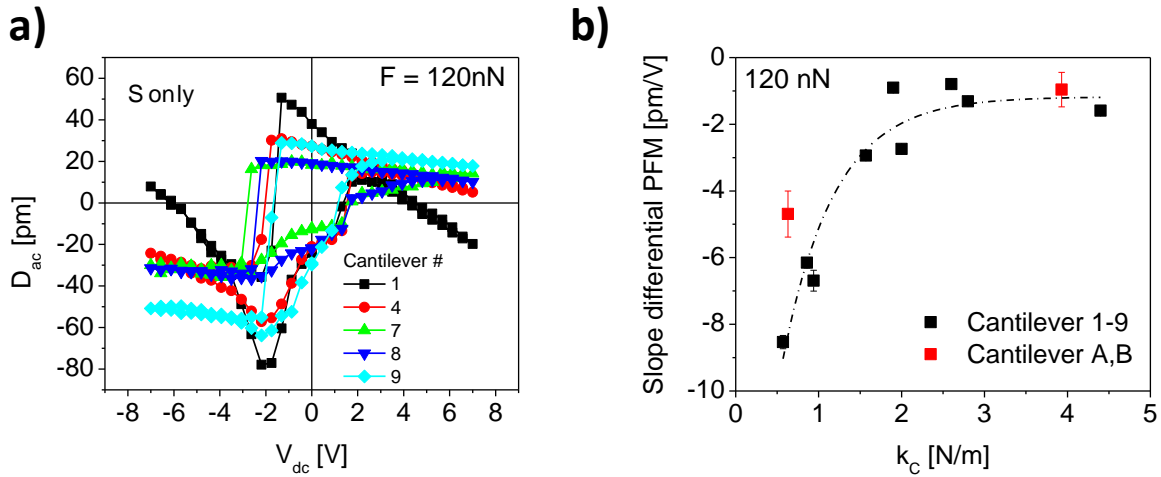

**Figure 5.** (a) On-field PFM hysteresis loops for cantilevers with different stiffness. (b) Slope of the PFM curves extracted from the positive hysteresis branch as a function of the cantilever stiffness. Only the static sensitivity $S$ was used to perform the quantification of $D_{ac}$.

In a second example, a different set of PFM measurements was performed on a 50 nm thick Pb(Zr,Ti)$O_3$/(La,Sr)MnO$_3$ film grown epitaxially with pulsed laser deposition on a SrTiO$_3$ (001) substrate. As before, we employed SS-PFM [45] to measure on- and off-field PFM hysteresis loops. All experimental data for the PZT film were recorded at a single point. Due to the high quality and uniformity of the epitaxially grown film, significant lateral variations in sample behavior could not be observed, which allowed for single point measurements. The measurements were performed with a total of 9 different cantilevers as summarized in Table S2 of Supplementary data. As the first quantification step, only the static sensitivity *S* was used to express $D_{ac}$ in picometers. The on-field curves shown in figure 5a are hysteretic due to polarization switching. In addition, a tilt in the curves is observed due to the electrostatic contribution, which strongly depends on the probe. Figure 5b shows the slope as function of cantilever stiffness from this as well as the two cantilevers used for PPLN. In general, the slope is higher for lower stiffness but saturates for higher one.

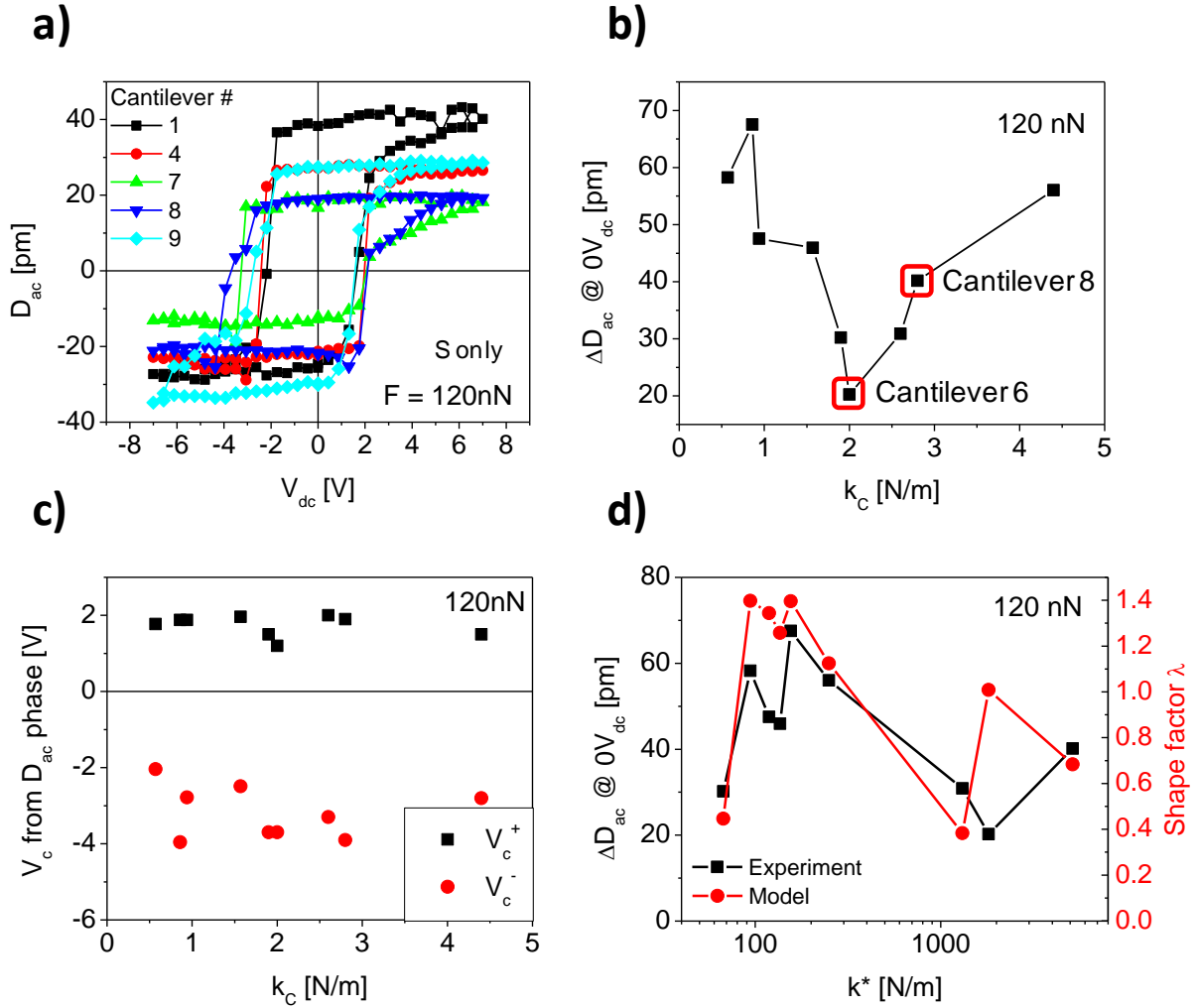

**Figure 6.** (a) Off-field PFM hysteresis loops for cantilevers with different stiffness. Cantilever static sensitivities *S* were used to convert the microscope signal into picometers for the plots. (b) Loop opening at 0 $V_{dc}$ and (c) coercive voltages $V_c^{+/-}$ extracted from the measured off-field PFM loops in (a) and shown versus cantilever stiffness. The data points for the two cantilevers, which are of the same model and manufacturer but differ in stiffness, are highlighted in (b). (d) Comparison of the measured loop opening from (b) and corresponding calculated shape factors shown versus

contact stiffness. The contact force was 120 nN in all the measurements. Cantilever properties are provided in Table S2 of Supplementary data.

In the case of the PZT thin film, we assume that the surface potential is not constant during the measurement since domain switching occurs. Therefore, we cannot apply the same combination of KPFM and SS-PFM approach as for the PPLN crystal. Instead, we turn our attention to the off-field PFM loops shown in figure 6a. During the off-field loop acquisition, $V_{ac}$ is applied after $V_{dc}$ voltage pulses, and electrostatic signal contributions are minimized. However, there is still a strong cantilever-dependence when only static sensitivity is used to extract the surface displacement in picometers. This again demonstrates that the static sensitivity of a probe cannot serve to correctly predict the intensity of the cantilever response at the first-mode resonance in a PFM experiment.

The parameters extracted from the off-field loops to characterize the ferroelectric switching properties include the loop opening $\Delta D_{ac}$ at 0 $V_{dc}$ (related to switchable polarization of the ferroelectric film) and the coercive voltage $V_c$. These parameters are shown in figure 6b and c as function of a cantilever stiffness $k_C$ for all the probes. The coercive voltages were determined through the phase flips in the PFM phase loops. It can be seen that there is a non-monotonic trend for the loop opening without clear dependencies on any cantilever properties, including cantilever length, cantilever area, free resonance frequency, contact resonance frequency and $Q$ factor. From the experimental data, it is obviously impossible to compare directly two different measurements with two different cantilevers, even if they are nominally identical (the same model and from the same manufacturer: compare cantilevers 6 and 8 highlighted in figure 6b). In turn, the coercive voltages (figure 6c) do not seem to depend on the cantilever choice.

As described above, for each cantilever the tip-sample contact stiffness $k^*$ was determined from experimental values of the contact resonance frequencies and cantilever geometry. From this, the shape factor $\lambda$ was calculated for every cantilever (figure 6d) and compared with the experimentally measured ferroelectric loop opening plotted versus contact stiffness $k^*$. As evident, there is a very good match in the trends of the measured PFM loop opening and the calculated shape factors validating both the model and the data processing procedure. There is no exact correlation between data and theory, which is unsurprising considering the use of an idealized model. Deviations can be explained by a number of factors, such as inaccuracies in determination of cantilever parameters, *e. g.,* length and pyramid height and area, as well as imperfections of the cantilever shape. The actual tip displacement is correctly measured for a shape factor of unity, and it is approximately 53 pm in the experiment. The presence of electrostatic signal contributions shifts the piezoelectric loop along the vertical axis but does not change the loop opening (see also figure 2d). If we assume that the surface potential is constant, the piezoelectric constant $d$ can be calculated from the loop opening, which yields $d = 13.25$ pm/V. Unlike the PPLN case, here we do not compare the results to (unknown) macroscopic value and rather consider this to be a calibrated locally measured response. We further note that the position of the laser spot along the cantilever length may change the results as well. This was not investigated in this work where the laser spot was aligned the same way (above the sensing tip) for all cantilevers.

**Conclusions**

In conclusion, contact-resonance-enhanced AFM techniques used to study electrical and mechanical materials properties on the nanoscale can be more quantitative by taking into account the cantilever beam shape in the first eigenmode, which is a strong function of the tip-sample contact stiffness. Here, we have introduced a method to quantify surface displacements with a high accuracy based on an analytical model of the AFM cantilever vibrations. The proposed method was verified with two different high-quality well-characterized ferroelectric materials and a variety of cantilevers through measurements of piezoelectric constant in the resonance-enhanced piezoresponse force microscopy in a controlled environment. First, we measured nanometer-scale piezoelectric constant of a commercial periodically poled lithium niobate crystal and found an excellent match with the known macroscopic value specified by the sample manufacturer. In these experiments, we have for the first time introduced a methodology to quantitatively measure piezoelectric constants independently from cantilever properties and other factors in the PFM measurements. Second, the technique was verified with use of a well-defined ferroelectric film. The comparison of the variations in the hysteresis loop opening measured at varying contact stiffness with a number of cantilevers showed excellent agreement with the calculated factors for the cantilever shape correction. We find that the shape correction factors are strongly dependent on the cantilever choice, and for certain cantilevers, they remain close to unity in a broad range of contact stiffness. This is important for the simulation-guided cantilever choice tailored for dynamic AFM-based characterization techniques. From a broader perspective, these results are of a significant value for quantitative measurements with a range of SPM-based techniques.

**Acknowledgements**

Experiments were planned and conducted through personal support provided by the U.S. Department of Energy, Basic Energy Sciences, Materials Sciences and Engineering Division through the Office of Science Early Career Research Program (N.B.). The facilities to perform the experiments were provided at the Center for Nanophase Materials Sciences, which is sponsored at Oak Ridge National Laboratory by the Scientific User Facilities Division, Office of Basic Energy Sciences, U.S. Department of Energy, which also provided additional personal support (S. J., B. C., S.V.K., A.T.). P.Y. provided the ferroelectric PZT sample with the supports from the National Basic Research Program of China (Grant No. 2015CB921700) and National Natural Science Foundation of China (Grand No. 11274194).

# Supplementary data

# Quantification of surface displacements and electromechanical phenomena via dynamic atomic force microscopy

Nina Balke[1*], Stephen Jesse[1], Pu Yu[2,3,4], Ben Carmichael[5], Sergei V. Kalinin[1] and Alexander Tselev[1]

[1]Center for Nanophase Materials Sciences, Oak Ridge National Laboratory, Oak Ridge, TN 37831 United States

[2]State Key Laboratory for Low-Dimensional Quantum Physics, Department of Physics, Tsinghua University, Beijing, China

[3]Collaborative Innovation Center of Quantum Matter, Beijing, China

[4]RIKEN Center for Emergent Matter Science (CEMS), Wako, Saitama 351-0198, Japan

[5]Southern Research, Birmingham, AL 35211, United States

*Corresponding author: balken@ornl.gov

The Supplementary data provide a more detailed analysis and characterization of the cantilevers as well as the measured ferroelectric loops for all cantilevers used in this study. In addition, expressions for the solution for the cantilever harmonic oscillations that were used in the paper are presented.

**Characterization of the cantilevers used in the experiments with a PPLN crystal**

**Table S1.** Cantilever properties: length ($L$), width ($W$), thickness ($T$), free resonance frequency ($f_0$), contact resonance frequency ($f_c$), cantilever spring constant ($k_C$), contact stiffness ($k^*$), inverse of the static cantilever sensitivity ($S = (s \cdot s'_{static})^{-1}$), $L$, $W$, and $T$ are according to the manufacturer specifications. $f_c$ and $k^*$ are experimentally determined with a 120 nN tip-sample contact force.

| Cantilever | L [μm] | W [μm] | T [μm] | $f_0$ [kHz] | $f_c$ [kHz] | $k_C$ [N/m] | $k^*$ [N/m] | S [nm/V] |
|---|---|---|---|---|---|---|---|---|
| A | 225 | 28 | 2 | 78.8 | 382 | 3.9 | 380 | 98 |
| B | 250 | 35 | 2 | 34.6 | 185 | 0.63 | 145 | 107 |

**Characterization of the cantilevers used for the experiments with a PZT epitaxial film**

**Table S2.** Cantilever properties: length ($L$), width ($W$), thickness ($T$), free resonance frequency ($f_0$), contact resonance frequency ($f_c$), cantilever spring constant ($k_C$), contact stiffness ($k^*$), inverse of the static cantilever sensitivity ($S = (s \cdot s'_{static})^{-1}$). $L$, $W$, and $T$ are given according to the manufacturer specifications. Free resonance frequencies marked with * are nominal and not measured. $f_c$ and $k^*$ are experimentally determined with a 120 nN tip-sample contact force. The cantilevers are ordered in the table according to the ascending cantilever stiffness $k_C$.

| Cantilever | L [μm] | W [μm] | T [μm] | $f_0$ [kHz] | $f_c$ [kHz] | $k_C$ [N/m] | $k^*$ [N/m] | S [nm/V] |
|---|---|---|---|---|---|---|---|---|
| 1 | 350 | 35 | 2 | 20 | 99 | 0.57 | 94 | 148 |
| 2 | 300 | 35 | 2 | 27 | 136 | 0.86 | 156 | 132 |
| 3 | 300 | 35 | 2 | 28 | 135 | 0.94 | 119 | 132 |
| 4 | 250 | 35 | 2 | 39 | 191 | 1.6 | 136 | 104 |
| 5 | 90 | 32.5 | 1 | 130* | 702 | 1.9 | 67 | 42 |
| 6 | 250 | 35 | 2 | 40* | 249 | 2 | 1816 | 115 |
| 7 | 110 | 32.5 | 1 | 90* | 710 | 2.6 | 1309 | 68 |
| 8 | 250 | 35 | 2 | 40* | 271 | 2.8 | 5149 | 128 |
| 9 | 225 | 28 | 3 | 75* | 339 | 4.4 | 250 | 102 |

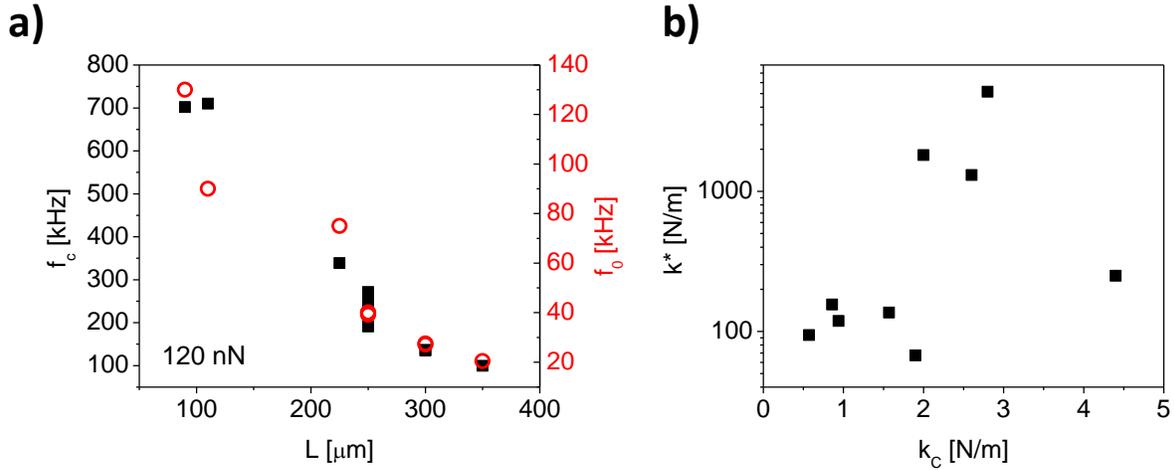

**Figure S1.** Cantilever properties (contact resonance frequency and contact stiffness) and their relation to $L$ and $k_C$.

**Additional data obtained with the PZT epitaxial film**

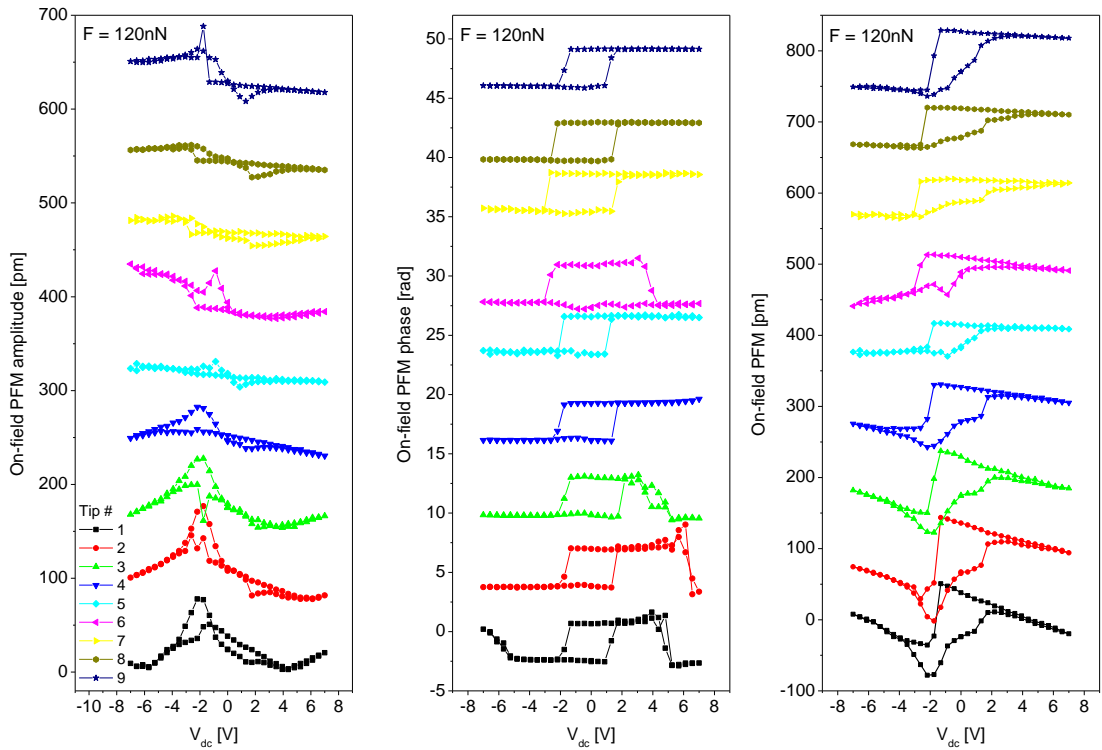

**Figure S2.** On-field PFM amplitude, phase, and mixed signal loops for all the cantilevers in Table S2 sorted from lowest (bottom) to highest (top) cantilever stiffness. All measurements were carried out with a 120 nN tip-sample contact force.

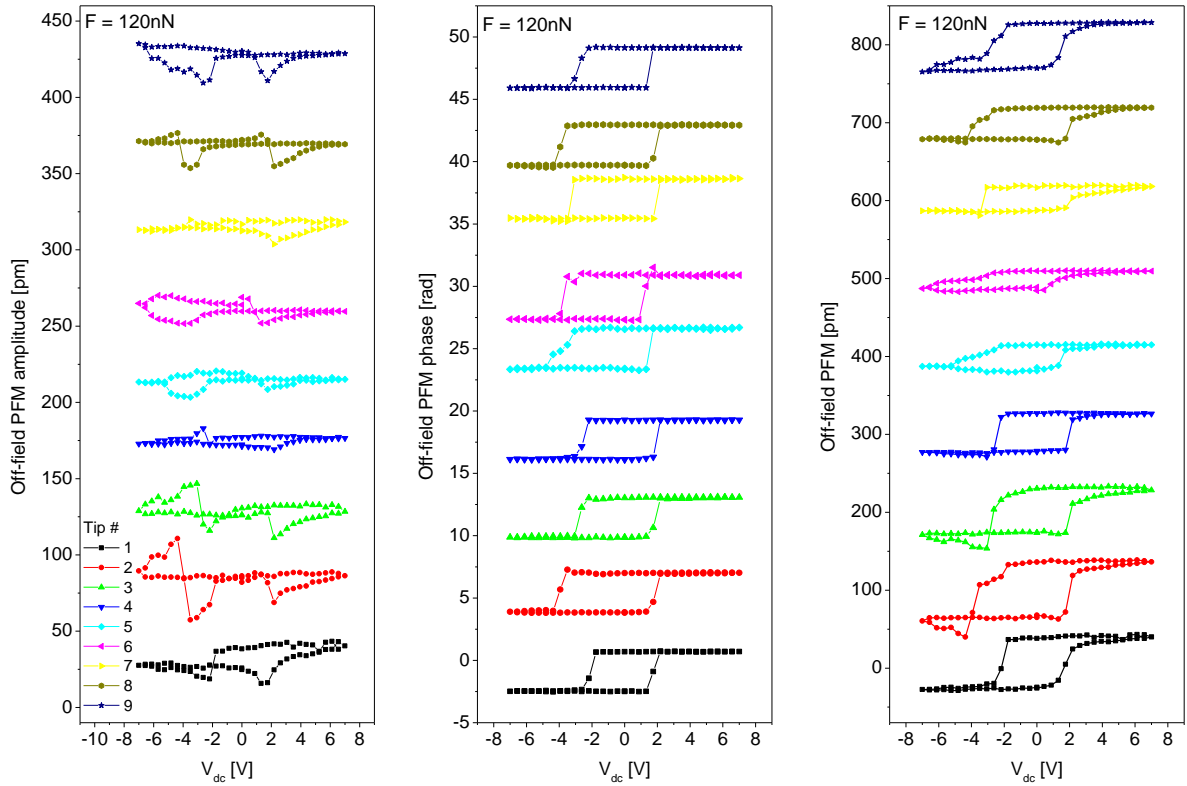

**Figure S3.** Off-field PFM amplitude, phase, and mixed signal loops for all the cantilevers in Table S2 sorted from lowest (bottom) to highest (top) cantilever stiffness. All measurements were carried out with a 120 nN tip-sample contact force.

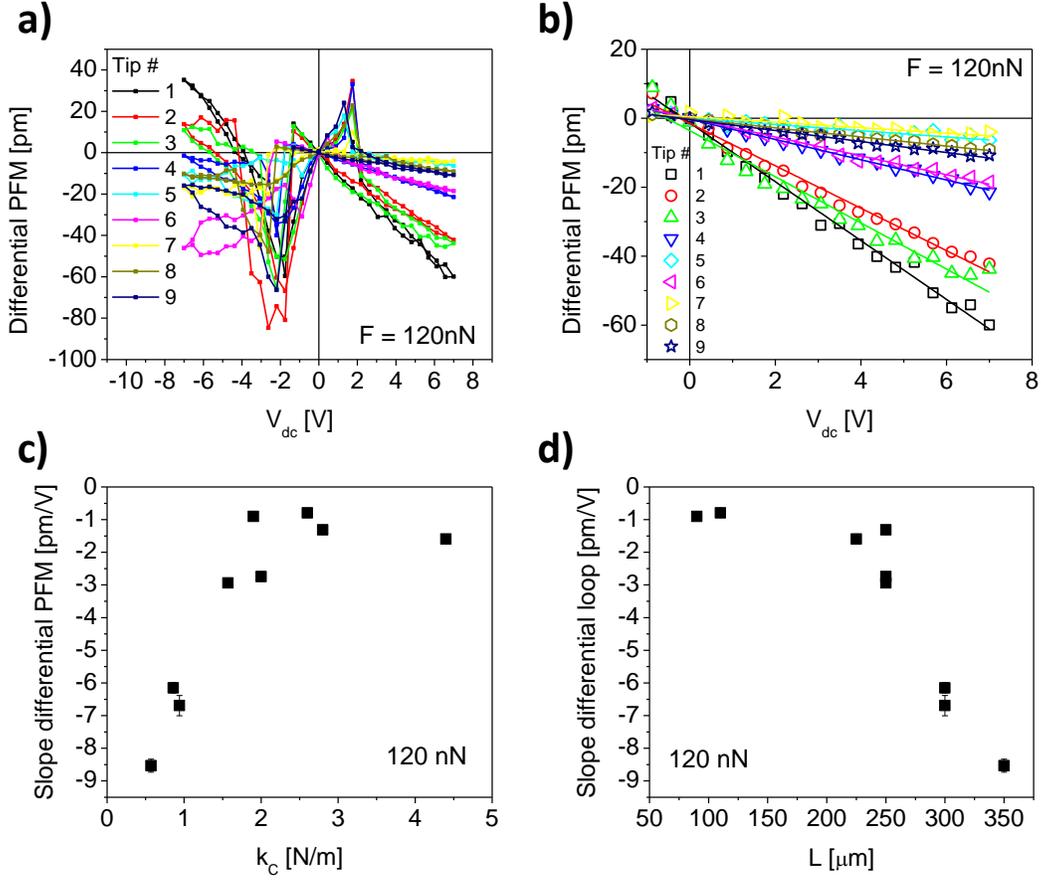

**Figure S4.** Differential PFM loops for all the cantilevers in Table S2: (a) full voltage range and (b) zoomed into the positive voltage branch where no domain switching occurs. Fitted slope of the curves shown in (b) as a functions of (c) cantilever stiffness and (d) cantilever length.

**Analytical solution for cantilever harmonic oscillations**

The provided expressions are adapted from U. Rabe, "Atomic force acoustic microscopy," *Applied scanning probe methods II*, B. Bhushan and H. Fuchs (Editors), Springer Berlin Heidelberg, 2006, pp. 37-90. Refer to figure 1a in the main text for notations.

Solutions of the cantilever vibration equation:

$$y(x) = A_2(\cos \alpha x - \cosh \alpha x) + A_4(\sin \alpha x - \sinh \alpha x) \quad (S.1)$$

$$y_2(x_2) = A_1(\cos \alpha x_2 + \cosh \alpha x_2) + A_3(\sin \alpha x_2 + \sinh \alpha x_2)$$

Dispersion relation:

$$\alpha = \frac{1}{L} 1.8751 \left( \left(\frac{f}{f_{1,\text{free}}}\right)^2 - i \frac{1}{Q_{1,\text{free}}} \frac{f}{f_{1,\text{free}}} \right)^{1/4} \quad (S.2)$$

where $f$ is frequency, and $f_{1,\text{free}}$ is the frequency of the first eigenmode of the free cantilever vibrations ($f_{1,\text{free}} \equiv f_0$ in Tables S1 and S2). The quality factor of the free cantilever vibrations $Q_{1,\text{free}}$ was fixed equal to 220 in the calculations described in the main text. Varying this value in a range from 120 to 300 did not influence the calculation results significantly.

Slope at the pyramid (sensing tip):

$$\theta = \left.\frac{dy}{dx}\right|_{x=L_1} = \left.\frac{dy_2}{dx_2}\right|_{x_2=L_2} \quad (S.3)$$

Notation used in the expressions for coefficients $A_1$-$A_4$:

$$s_1^\pm = \sin \alpha L_1 \pm \sinh \alpha L_1$$

$$s_2^\pm = \sin \alpha L_2 \pm \sinh \alpha L_2$$

$$c_1^\pm = \cos \alpha L_1 \pm \cosh \alpha L_1$$

$$c_2^\pm = \cos \alpha L_2 \pm \cosh \alpha L_2$$

$$ssh_1 = \sin \alpha L_1 \sinh \alpha L_1$$

$$ssh_2 = \sin \alpha L_2 \sinh \alpha L_2 \quad (S.4)$$

$$cch_1 = \cos \alpha L_1 \cosh \alpha L_1$$

$$cch_2 = \cos \alpha L_2 \cosh \alpha L_2$$

$$cch_1^\pm = 1 \pm \cos \alpha L_1 \cosh \alpha L_1$$

$$cch_2^\pm = 1 \pm \cos \alpha L_2 \cosh \alpha L_2$$

$$cch_{1+2}^\pm = 1 \pm \cos \alpha(L_1 + L_2) \cosh \alpha(L_1 + L_2)$$

$$mix_1^\pm = \sin\alpha L_1 \cosh\alpha L_1 \pm \cos\alpha L_1 \sinh\alpha L_1$$

$$mix_2^\pm = \sin\alpha L_2 \cosh\alpha L_2 \pm \cos\alpha L_2 \sinh\alpha L_2$$

Contact functions:

$$\phi(\alpha) = 3\frac{k^*}{k_C} + i(\alpha L_1)^2 p$$

$$p = \frac{L_1}{L}\frac{3\gamma\omega_0}{(1.8751)^2 k_C}$$

$$\phi_{Lon}(\alpha) = 3\frac{k^*_{Lon}}{k_C} + i(\alpha L_1)^2 p_{Lon}$$

$$p_{Lon} = \frac{L_1}{L}\frac{3\gamma_{Lon}\omega_0}{(1.8751)^2 k_C}$$

(S.5)

Auxiliary functions:

$$T(\alpha) = \frac{h^2}{L_1^3}\phi(\alpha)\sin^2\varphi + \frac{h^2}{L_1^3}\phi_{Lon}(\alpha)\cos^2\varphi$$

$$X(\alpha) = \frac{h}{L_1^3}\sin\varphi\cdot\cos\varphi\cdot[\phi_{Lon}(\alpha) - \phi(\alpha)]$$

(S.6)

$$U(\alpha) = \frac{1}{L_1^3}\phi(\alpha)\cos^2\varphi + \frac{1}{L_1^3}\phi_{Lon}(\alpha)\sin^2\varphi$$

Denominator:

$$N(\alpha) = 2\{-2\alpha^5 cch_{1+2}^+ + \alpha^4 T(mix_2^+ cch_1^- - cch_2^+ mix_1^+) - 2\alpha^3 X(cch_2^+ ssh_1 + ssh_2 cch_1^-)$$
$$+ \alpha^2 U(mix_2^- cch_1^- - cch_2^+ mix_1^-) - \alpha(TU - X^2)cch_2^+ cch_1^-\}$$

(S.7)

Characteristic equation:

$$N(\alpha) = 0 \tag{S.8}$$

Coefficients $A_1$-$A_4$ (with correction of misprints in the original paper):

$$A_1 = \{-\alpha^3 a_0[c_2^+ ssh_1 + s_2^+ mix_1^+ - c_2^- cch_1^-] - \alpha^2 b_0[c_2^+ mix_1^- + s_2^+ ssh_1 - s_2^- cch_1^-]$$
$$+ \alpha(Xa_0 - Tb_0)c_2^+ cch_1^- + (Xb_0 - Ua_0)s_2^+ cch_1^-\}/N(\alpha)$$

(S.9)

$$A_2 = \{\alpha^3 a_0[mix_2^+ s_1^- + ssh_2 c_1^- + cch_2^+ c_1^+] + \alpha^2 b_0[mix_2^- c_1^- + ssh_2 s_1^- + cch_2^+ s_1^+]$$
$$+ \alpha(Xa_0 - Tb_0)cch_2^+ c_1^- - (Xb_0 - Ua_0)cch_2^+ s_1^-\}/N(\alpha)$$

$$A_3 = \{\alpha^3 a_0[c_2^+ mix_1^+ - s_2^- ssh_1 + s_2^+ cch_1^-] - \alpha^2 b_0[-c_2^+ ssh_1 + s_2^- mix_1^- + c_2^- cch_1^-]$$
$$+ \alpha(Xa_0 - Tb_0)s_2^- cch_1^- - (Xb_0 - Ua_0)c_2^+ cch_1^-\}/N(\alpha)$$

$$A_4 = \{-\alpha^3 a_0[-ssh_2 s_1^+ + mix_2^+ c_1^- - cch_2^+ s_1^-] + \alpha^2 b_0[-ssh_2 c_1^- + mix_2^- s_1^+ - cch_2^+ c_1^+]$$
$$+ \alpha(Xa_0 - Tb_0)cch_2^+ s_1^+ + (Xb_0 - Ua_0)cch_2^+ c_1^-\}/N(\alpha)$$

with

$$a_0 = -u_0 \frac{h}{L_1^3} \phi_{Lon}(\alpha) \sin\varphi$$

$$b_0 = u_0 \frac{1}{L_1^3} \phi(\alpha) \cos\varphi \qquad (S.10)$$

$$\omega_0 = 2\pi f_{1,\text{free}}$$